# AI+CAD Data Representation Architecture: From DeepCAD Solid Modeling to WHUCAD Industrial-Level Parametric Feature Modeling

Rubin Fan [1,2)], Fazhi He [1,2,3,4,5)*], Yuxin Liu [2)], Jing Lin [3)], Ruibo Wan [3)], Xuecheng Zhang [1)], Qingchen Kong [1)]

1) School of Artificial Intelligence, Wuhan University, Wuhan 430072

2) School of Computer Science / Characteristic Demonstration Software College, Wuhan University, Wuhan 430072

3) School of Cyber Science and Engineering, Wuhan University, Wuhan 430072

4) National College for Excellent Engineers, Wuhan University, Wuhan 430072

5) National Engineering Research Center for Multimedia Software, Wuhan University, Wuhan 430072

## Abstract

In July 2025, Study Times, sponsored by the Party School of the Central Committee of the CPC, pointed out that 95% of industrial software for R&D and design in China relies on imports, and that 90% of the high-end CAD/CAE/CAM software market is monopolized by European and American giants. This is a typical strategic bottleneck problem. Unlike the visually oriented goal of "visual plausibility" pursued by related sister disciplines such as CV and CG, CAD places greater emphasis on "industrial usability". In CAD, data representation architecture is more foundational than the optimization of network algorithms. This paper first starts from data representation in AI+CAD and reports a classification paradigm and research progress in AI+CAD. Then, using the open-source DeepCAD data representation as an example, it analyzes the pain points of representative AI+CAD work and the gap between such work and real industrial-level parametric feature modeling. Next, by comparison with the open-source WHUCAD data representation, it discusses how its three-level architecture provides fundamental support for industrial-level parametric feature modeling. Finally, in view of the rapid iteration of the AI wave, large models, and agents, this paper offers an outlook on AI+industrial-level CAD.

## Introduction

At present, deep-learning-based three-dimensional modeling methods have generally reached the stage of "solid modeling". However, there remains an obvious generational gap and a deep divide between these methods and the "design-history-based parametric feature modeling methods" actually used in high-end industrial CAD systems today. In this paper, high-end industrial CAD systems refer to strategically critical systems represented by CATIA. This gap has become a major pain point in the intelligent development of CAD industrial software. AI+CAD represented by DeepCAD has fundamental defects in the underlying architecture of its data representation. No matter how much the existing dataset size is expanded, or how advanced the machine-learning method is - for example large models, multimodal models, diffusion models, or Mamba - it is difficult to overcome the above gap. Taking the open-source DeepCAD project and the open-source WHUCAD project as examples, this paper compares and interprets the above gap and pain points, hoping to make a modest contribution to helping China's AI+CAD industrial software avoid falling once again into a strategic bottleneck in this new track.

The essence of deep learning is the learning of data representations. In Chinese, "data representation" is often used to refer to the high-dimensional features extracted and organized by a network model from input data during training. By contrast, "data representation format" emphasizes the structured organization of data before it enters the network model: that is, in what form the research object is encoded, and which semantic relationships and engineering information are retained. For tasks involving images, text, or discrete geometry, data representation formats are relatively stable, and the focus of deep learning lies more in optimizing network architectures and feature-representation learning algorithms. For CAD, however, the data representation format itself is the fundamental issue that determines whether AI can enter industrial workflows. The same CAD model may be represented as a point cloud, a mesh, a boundary representation model, or as sketches, constraints, a feature tree, modeling order, and the mechanisms of topological naming and selection. Different representation formats preserve completely different engineering semantics and directly determine the upper bound of what a network model can learn. Therefore, this paper adopts the foundational concept of data representation format and focuses on how data objects are organized and represented in AI+CAD, rather than on internal neural feature representations or network algorithm optimization.

The structure of the paper is as follows. Section 1 starts from the data representation of AI+CAD and reports a classification of AI+CAD paradigms and their research progress. Section 2 uses DeepCAD data representation as an example to analyze the pain points of AI+simple CAD. Section 3 compares and interprets the WHUCAD data representation and reports how its three-level data representation architecture supports AI+industrial-level parametric CAD at the foundational level. Section 4 concludes the paper.

## 1 AI+CAD Paradigms: A Classification and Progress Based on Data Representation

Unlike the visually oriented goal of "visual plausibility" pursued by related sister disciplines such as CV and CG, CAD places greater emphasis on "industrial usability". Data representation is more foundational than network algorithms. The data representation of AIGC and the industrial usability of generated results constitute a gold standard for evaluating AI+CAD paradigms. Therefore, this paper uses the data representation and generated results of AIGC as a clue for classifying and reviewing the progress of existing AI+CAD research, as shown in Figure 1.

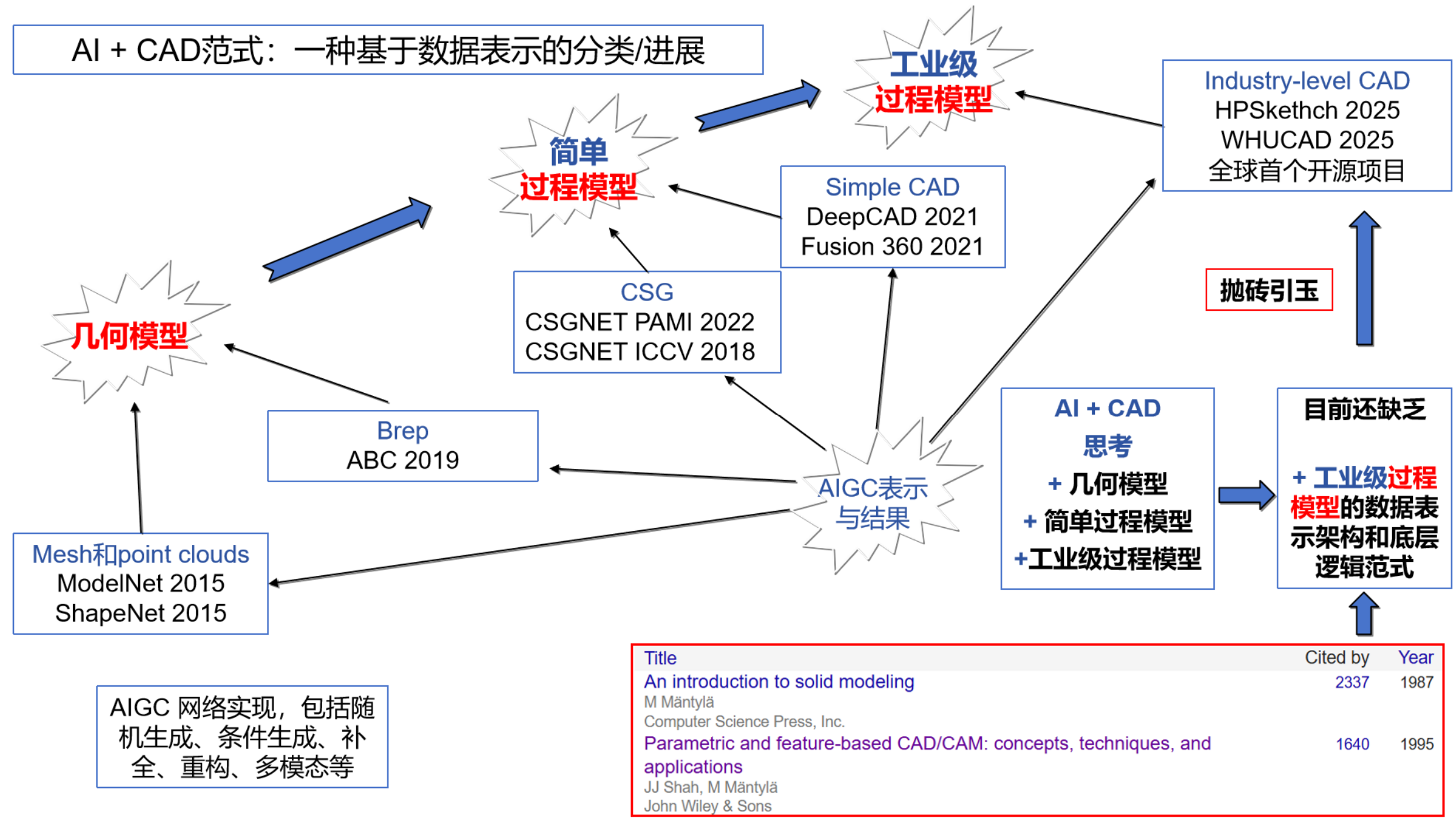

Figure 1. AI+CAD paradigms: a classification and progress study based on data representation

## 1.1 Paradigm 1: AI + Geometric Representation

In deep learning research in computer graphics and three-dimensional computer vision, data representation based on three-dimensional geometry is fundamental. This paradigm takes geometric shape as the main learning object. Typical data representations include voxels, point clouds, triangular meshes, implicit fields, and boundary representation models.

ModelNet, a representative geometric dataset using discrete and approximate representations, claims that "We construct ModelNet - a large-scale 3D CAD model dataset." In reality, however, it provides only triangular meshes. ShapeNet claims that "It is a richly annotated, large-scale repository of shapes represented by 3D CAD models of objects," but it provides only point clouds and triangular meshes. These two types of data representation mainly serve 3D visual recognition, reconstruction, and generation. They focus on object appearance and semantic categories rather than design history, constraint relationships, or more complex industrial-level parametric feature trees [1-2]. Therefore, even if AIGC systems based on such data representations can generate visually plausible three-dimensional shapes, they usually cannot directly enter engineering design workflows that require parameter modification and feature reuse.

The ABC dataset claims in its title to be "A Big CAD Model Dataset For Geometric Deep

Learning", but Section 3.1 further explains that "We will use the term CAD model to refer to a geometric shape defined by a boundary representation" [3]. Such geometric representations push the research object further toward continuous and exact boundary representation models, enabling geometric deep learning to better exploit analytical curves, surface patches, normals, curvature, and other geometric attributes. Subsequently, work such as BRepGen began to attempt direct generation of boundary representation models, indicating that AI+CAD research has gradually moved from discrete and approximate representations toward continuous and exact boundary representation models used by CAD kernels [4]. The important contribution of this class of data representation is that it lifts shapes from being merely "visually plausible" to being "geometrically more accurate" boundary representation models. Geometry learning networks based on boundary representation, such as BRepNet, ParSeNet, PrimitiveNet, and SolidGen, are not discussed further here [5-9].

AIGC based on geometric representations such as point clouds, meshes, and boundary representations can only solve the question of "what" is generated. It cannot solve the question of "how engineers model step by step in the actual modeling process." Therefore, when AI directly generates geometric models, there are at least two industrial-usability problems.

Problem 1. As Piegl, professor at the University of South Florida and former editor-in-chief of Computer-Aided Design, pointed out in the top ten challenges in CAD: "While one can cheat the eye in computer graphics and animation, the milling machine is not as forgiving" [10]. In other words, in computer graphics and animation, geometric errors may be visually imperceptible and can "deceive the eye"; however, in machining and manufacturing, machine tools do not tolerate geometric errors. Therefore, even if AI+geometric models can generate visually plausible three-dimensional shapes, they cannot directly guarantee that geometric accuracy, topological validity, and engineering manufacturability meet the requirements of industrial CAD scenarios.

Problem 2 concerns parameterization, editability, and reusability, which have also motivated recent research on AI+procedural models. Industrial CAD models need not only describe final geometry; they also need to preserve modeling history, sketch constraints, parameter associations, and industrial-level parametric feature trees, so as to support subsequent dimensional modification and feature reuse. Thus, merely generating geometric models is insufficient for industrial-level parametric design, and procedural models must be further adopted. Procedural models call the CAD geometry core/kernel to output final geometric models. This can, to some extent, avoid or alleviate precision problems associated with directly generated geometric models, while preserving the parametric modeling process and editable structure, making the representation closer to what industrial CAD software truly requires.

### 1.2 Paradigm 2: AI + Simple Procedural Representation

Compared with Paradigm 1, Paradigm 2 takes one step forward by representing the CAD modeling process as a simple sequence of CAD commands. What the AI network learns and outputs is not an explicit geometric representation, but an implicit procedural representation. For machine learning, this serialized expression is highly friendly and facilitates the use of Transformers, diffusion models, autoregressive models, and multimodal conditional generation frameworks.

DeepCAD is representative of this paradigm. It represents three-dimensional models as CAD operation token sequences similar to natural language and uses a Transformer for autoencoding and random generation. Its open dataset contains 178,238 CAD models and their corresponding modeling sequences [11]. Fusion 360 Gallery Reconstruction adopts a similar data representation and provides 8,625 design sequences [12]. There are many learning networks and optimization works based on AI+simple CAD procedural models, which are not discussed further here [13][14].

In Paradigm 2, CSG-Net adopts the representation of constructive solid geometry (CSG). Compared with the data representations of DeepCAD and Fusion 360, its underlying logic is similar [15]. DeepCAD command sequences first generate geometric solid primitives through simple sketch extrusion, and then finally obtain geometric results through a series of Boolean operations. The difference is that CSG directly provides parametric geometric solid primitives - for example, directly using length, width, and height to represent a cube - whereas DeepCAD gives parametric geometric solid primitives through sketch extrusion - for example, first using a parametric sketch quadrilateral and then an extrusion-height parameter to represent a cube. In essence, both belong to the category of solid modeling and support only primitive-level parameterization.

With reference to two milestone monographs in CAD [16][17], Paradigm 2 is essentially AI+solid modeling plus a geometry kernel.

If benchmarked against the level of industrial CAD technology in the 1980s and 1990s, this AI+simple procedural representation, that is AI+solid modeling, is already a closed loop. It can support solid modeling and directly interface with geometry kernels. Almost all methods following DeepCAD directly connect to the OCC kernel and obtain geometric results through Boolean operations with global parameters.

However, after many years of development, contemporary industrial-level CAD systems - in this paper, high-end industrial CAD systems represented by CATIA - have evolved from "solid modeling" to "design-history-based parametric feature modeling." Therefore, Paradigm 2 cannot support feature modeling in contemporary industrial-level CAD.

### 1.3 Paradigm 3: AI + Native Industrial-Level Procedural Representation

As preliminary and exploratory work, the open-source projects HPSketch and WHUCAD construct industrial-level CAD data representations for two-dimensional sketches and three-dimensional modeling. The dataset, deep-learning code, and data foundation are publicly available, and the data foundation is free to use. Unlike previous paradigms, the industrial-level CAD procedural representation of WHUCAD does not represent merely a solid whose shape "looks like a part". Instead, it represents a parametric feature model that can continue to be solved, edited, reused, and exported by a CAD system. Therefore, the WHUCAD data representation unifies design history, sketch constraints, feature operations, topological element naming, and selection into the same data representation architecture [18][19].

Thus, the significance of HPSketch and WHUCAD is not merely the addition of more CAD command types. Rather, they upgrade AI+CAD data representation from "simple geometric operation commands" to "feature operation commands". Features are a class of ontological concepts and rule systems that express design semantics and manufacturing processes in modern CAD/CAM systems.

Taking the chamfer feature in CAD design semantics as an example, solid modeling can obtain a chamfer shape by specifying global parameters for Boolean operations. Such methods were already used in solid modeling in the 1980s, but they are difficult to edit and reuse, and in actual industrial deployment they incur huge manual interaction costs. In modern industrial-level parametric feature CAD systems, however, the underlying logic and ontological concepts of the chamfer feature are as follows.

- In general, a feature is both a high-level CAD operation command expressing an engineer's design semantics and the geometric result of a feature operation, such as an edge feature or face feature.
- For a chamfer feature, in addition to the size parameter of the chamfer, this high-level operation must be associated with a solid edge. The edge represents the geometric object to be modified in the engineer's design semantics; it may also be called an edge feature.
- This geometric object, namely the edge feature, is itself the result of a preceding series of historical feature operations. The edge itself is a feature. In addition to having its own geometric shape - the geometric shape that solid modeling can represent - the edge feature logically contains all modeling history up to the current point and can be traced all the way back to the first operation.
- The modeling history, topological information, and geometric information together construct the native permanent naming of this topological element. In this way, the "chamfer feature" based on the "edge feature" can truly realize design-history-based parametric feature modeling, as described in Function 2 below.

Different from the single function of solid modeling, parametric feature modeling has the following two functions.

- Function 1: Through a human-computer interactive CAD sequence process, it supports the construction of the current shape. This process is equivalent to solid modeling. The key difference is that, while constructing the current shape, it simultaneously constructs various features, including visible shape features and the feature dependencies behind visualization, including native permanent naming of topological elements, thereby supporting the subsequent Function 2.
- Function 2: For an already constructed parametric feature model, if the user edits or modifies a feature parameter, the CAD system replays the modeling history according to feature dependencies and permanent topological-element naming, without human-computer interaction. It then selects and locates feature entities and automatically propagates parameters, thereby automatically solving the global Boolean-operation parameters used when calling the geometry kernel and finally obtaining a new geometric shape.

Therefore, the essence of WHUCAD is AI+industrial-level parametric feature modeling.

## 2 DeepCAD: Pain Points of Simple Procedural Representation

Taking DeepCAD as an example, its data representation consists of CAD operation commands based on "simple sketches" and "simple extrusion operations". Although it claims to be a complete design history, DeepCAD's procedural representation actually lacks more fundamental key content. Compared with industrial-level parametric feature modeling, it exhibits an obvious generational gap and several pain points.

Referring to the two milestone monographs in the history of CAD development and to the example of the chamfer feature concept in Section 1.3, the simple CAD procedural commands in DeepCAD are essentially simple sketch commands in two dimensions or simple Boolean operations in three dimensions. They do not yet reach the concept of "feature modeling" in true parametric feature modeling.

Compared with contemporary industrial-level parametric feature modeling systems, the generational gap and pain points of DeepCAD fall into the following three aspects.

### 2.1 Insufficiency of Two-Dimensional Sketch Representation

For two-dimensional sketches, DeepCAD is far from industrial-level CAD modeling in both primitive expression and constraint expression.

In terms of primitives, DeepCAD contains only simple primitives such as lines, arcs, and circles. It is difficult for it to support advanced sketch primitives that are crucial in industrial-level CAD, such as splines, parabolas, hyperbolas, and ellipses. Taking splines as an example, after decades of efforts by CAD experts and scholars from computer science, mathematics, mechanical engineering, and other fields, NURBS curves and surfaces developed on this basis have become the soul and essence of shape modeling in modern industrial-level parametric CAD systems.

In terms of constraints, DeepCAD sketch procedures have no constraints, and therefore cannot support industrial-level parametric modeling or subsequent parameter modification and design-intent maintenance. In industrial-level parametric CAD, a sketch is a solvable design structure composed of primitives, dimensions, and constraints such as coincidence, parallelism, perpendicularity, tangency, and symmetry.

In terms of two-dimensional feature dependencies, because the sketch procedure in DeepCAD lacks mechanisms for naming and selecting two-dimensional topological elements, it cannot establish various constraint relationships among two-dimensional primitives during modeling and thus cannot support contemporary industrial-level CAD modeling.

### 2.2 Lack of High-Level Three-Dimensional Modeling Commands at the Feature Level

DeepCAD's three-dimensional modeling commands are merely simple Boolean operations and are unable to express high-level feature operations in industrial-level CAD. The procedural representation of DeepCAD expresses the modeling process as simple sketches and simple extrusions. There are no dependencies among multiple "sketch + extrusion" operations. In subsequent editing, relationships among these operations must be established manually to satisfy design semantics.

Therefore, DeepCAD's procedural representation cannot support high-level feature operations such as revolve, chamfer, fillet, mirror, draft, shell, groove, and revolved groove. These high-level features are basic requirements in contemporary industrial-level parametric feature modeling.

For example, taking sketch and extrusion commands as an example, DeepCAD supports only extrusion in a vertical direction with a vertical height. In parametric feature modeling, however, extrusion can be performed up to another specified face, thereby expressing richer design semantics of the designer.

Only CAD commands based on permanent topological-element naming are CAD feature operation commands in the true sense. Only such commands can support various advanced forms of parametric feature modeling, including the above "extrude up to a face" operation.

### 2.3 Lack of Permanent Naming and Selection Mechanisms for Three-Dimensional Topological Elements

In terms of permanent naming and selection mechanisms for three-dimensional topological elements, industrial-level parametric CAD systems require feature commands to specify their referenced objects. See Section 1.3 for the underlying logic and ontological concept of the chamfer feature in modern industrial-level parametric feature CAD systems. Modeling history, topological information, and geometric information together construct the native permanent naming of the edge feature as a topological element. Only on this basis can a chamfer feature based on an edge feature truly realize design-history-based parametric feature modeling.

Because DeepCAD lacks permanent topological-element naming and selection mechanisms, there are no dependencies among its multiple "sketch + extrusion" operations, and it cannot support true feature-level operations. Even in the simplest sketch extrusion, it cannot support extrusion up to another face.

Through the above analysis, DeepCAD data representation is essentially AI+solid modeling. Its editability and reusability are far from meeting the requirements of contemporary industrial-level parametric feature modeling.

Recent related studies show that DeepCAD (170k samples) and OmniCAD (450k samples), which are based on simple procedural representations, suffer from saturation issues [20].

## 3 WHUCAD: A Data Representation Architecture for Industrial-Level Parametric Feature Modeling

Unlike the simple procedural representation of DeepCAD, the data representation of WHUCAD is divided into a three-level architecture that broadly supports contemporary industrial-level parametric feature modeling.

In the era of AI+CAD, the progress from DeepCAD to WHUCAD is, to a certain extent, equivalent to the progress of traditional CAD systems at the end of the last century from solid modeling to parametric feature modeling [16][17].

### 3.1 The Three-Level Architecture of the Two-Dimensional Sketch Dataset HPSketch

The WHUCAD two-dimensional sketch component, HPSketch, establishes a three-level architecture consisting of Level 0, Level 1, and Level 2. A sketch is no longer merely "which primitives have been drawn"; at each step of modeling, it further contains "which operations are performed on which primitives." Level 1 and Level 2 are original contributions. Even at Level 0, HPSketch supports industrial-level advanced primitives that are not supported in DeepCAD, CADL, or SketchGraphs.

Level 0 is the primitive-layer representation. At the primitive layer, HPSketch creates industrial-level advanced primitives such as ellipses, spline curves, parabolas, and hyperbolas.

The significance of this layer is that it improves the geometric expressive power of sketches, allowing network models to learn the soul and essence of real industrial-level CAD sketches rather than only simple lines, circles, and arcs.

Level 1 is the high-level feature-operation-layer representation. HPSketch creates advanced sketch data representations such as mirror, chamfer, fillet, and rotate. Such representations are absent from DeepCAD, CADL, and SketchGraphs. HPSketch is benchmarked against real industrial-level CAD sketch modeling. Engineers do not simply draw primitives one by one; they often construct complex contours, including constraints, quickly through the interaction pattern of "select existing primitives - execute high-level operation."

Level 2 is the selection-mechanism-layer representation and is the core component of the HPSketch three-level architecture. A key issue in high-level operations is whether the data representation can express "on which primitive, which group of primitives, or which contour loop an operation is performed." Therefore, HPSketch constructs naming rules and selection mechanisms for sketch primitives, enabling network models to learn the human interactive semantic logic of "select objects first, then perform operations." Level 2 provides reference objects for Level 1 and is the prerequisite for high-level operations to hold.

Figure 2 shows the architectural advantages of HPSketch compared with the sketches of DeepCAD, CADL, and SketchGraphs.

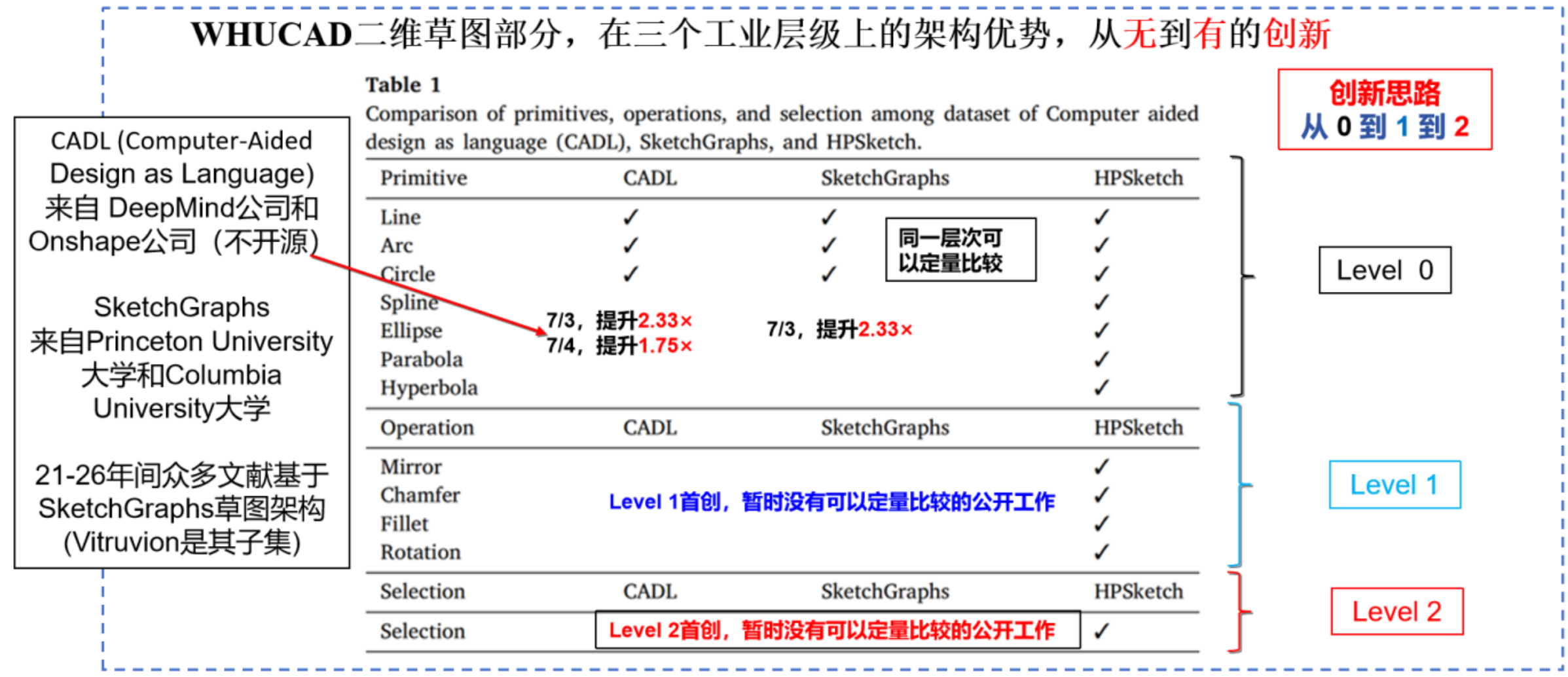

**Table 1**
Comparison of primitives, operations, and selection among dataset of Computer aided design as language (CADL), SketchGraphs, and HPSketch.

| Primitive | CADL | SketchGraphs | HPSketch |
|---|---|---|---|
| Line | ✓ | ✓ | ✓ |
| Arc | ✓ | ✓ | ✓ |
| Circle | ✓ | ✓ | ✓ |
| Spline | | | ✓ |
| Ellipse | | | ✓ |
| Parabola | | | ✓ |
| Hyperbola | | | ✓ |
| **Operation** | **CADL** | **SketchGraphs** | **HPSketch** |
| Mirror | | | ✓ |
| Chamfer | | | ✓ |
| Fillet | | | ✓ |
| Rotation | | | ✓ |
| **Selection** | **CADL** | **SketchGraphs** | **HPSketch** |
| Selection | | | ✓ |

Figure 2. Architectural advantages of the two-dimensional sketch component of WHUCAD

Further analysis and explanation beyond the architecture diagram are as follows.

First, sketches in SketchGraphs are actually graph representations of the result of "primitives + constraints", namely a geometric constraint graph. They do not support procedural modeling. Based on this representation, a network can only learn the result, but not the process or the reason. It is a typical case of "knowing what is so, but not knowing why it is so." By contrast, HPSketch supports procedural modeling and therefore knows not only what is so, but also why it is so.

Second, the constraints in HPSketch are based on "process + selection." The constraints beyond the above architecture diagram are shown in Figure 3, and overall they cover the sketch-constraint types used in mainstream flagship CAD parametric modeling systems.

| Constraint | ......(the former digits) | $\alpha_c$ | $l_1$ | $l_2$ | $t_c$ |
|---|---|---|---|---|---|
| Fix | | | | | 0 |
| Distance | | | | $l_2$ | 1 |
| Coincidence | | | | | 2 |
| Concentricity | | | | | 3 |
| Tangency | | | | | 4 |
| Length | | | | $l_2$ | 5 |
| Angle | | $\alpha_c$ | | | 6 |
| Parallelism | | | | | 7 |
| Horizontality | | | | | 8 |
| Perpendicularity | | | | | 9 |
| Verticality | | | | | 10 |
| Radius | | | $l_1$ | | 11 |
| Symmetry | | | | | 12 |
| Equidistance | | | | | 13 |
| MajorRadius | | | $l_1$ | | 14 |
| MinorRadius | | | $l_1$ | | 15 |

Figure 3. Constraint representation of HPSketch

### 3.2 The Three-Level Architecture of WHUCAD for Three-Dimensional Modeling

The three-dimensional modeling component of WHUCAD establishes a three-level architecture consisting of Level 0, Level 1, and Level 2. Level 1 and Level 2 are original contributions. Even at Level 0, WHUCAD supports advanced spline primitives that are absent from DeepCAD and Fusion 360.

At Level 1, WHUCAD creates high-level features in three-dimensional modeling representations, including revolve, groove, revolved groove, shell, chamfer, fillet, hole, mirror, and draft. Its Level 1 data representation broadly covers the high-level feature modeling capabilities of mainstream flagship CAD parametric modeling systems.

At Level 2, WHUCAD creates native permanent naming and selection mechanisms for three-dimensional topological elements. The selected objects in three-dimensional feature modeling may be edges or faces generated by preceding operations. For example, a chamfer operation must explicitly act on a particular edge, and a shell operation must explicitly remove or retain a particular face. WHUCAD has topological naming and selection mechanisms that establish reference relationships among different feature operations before and after a given operation, thereby supporting parameter propagation and feature reconstruction when the design history changes. Once preceding feature operations change, the design semantics can be automatically maintained. Figure 4 shows the architectural advantages of the three-dimensional modeling component of WHUCAD.

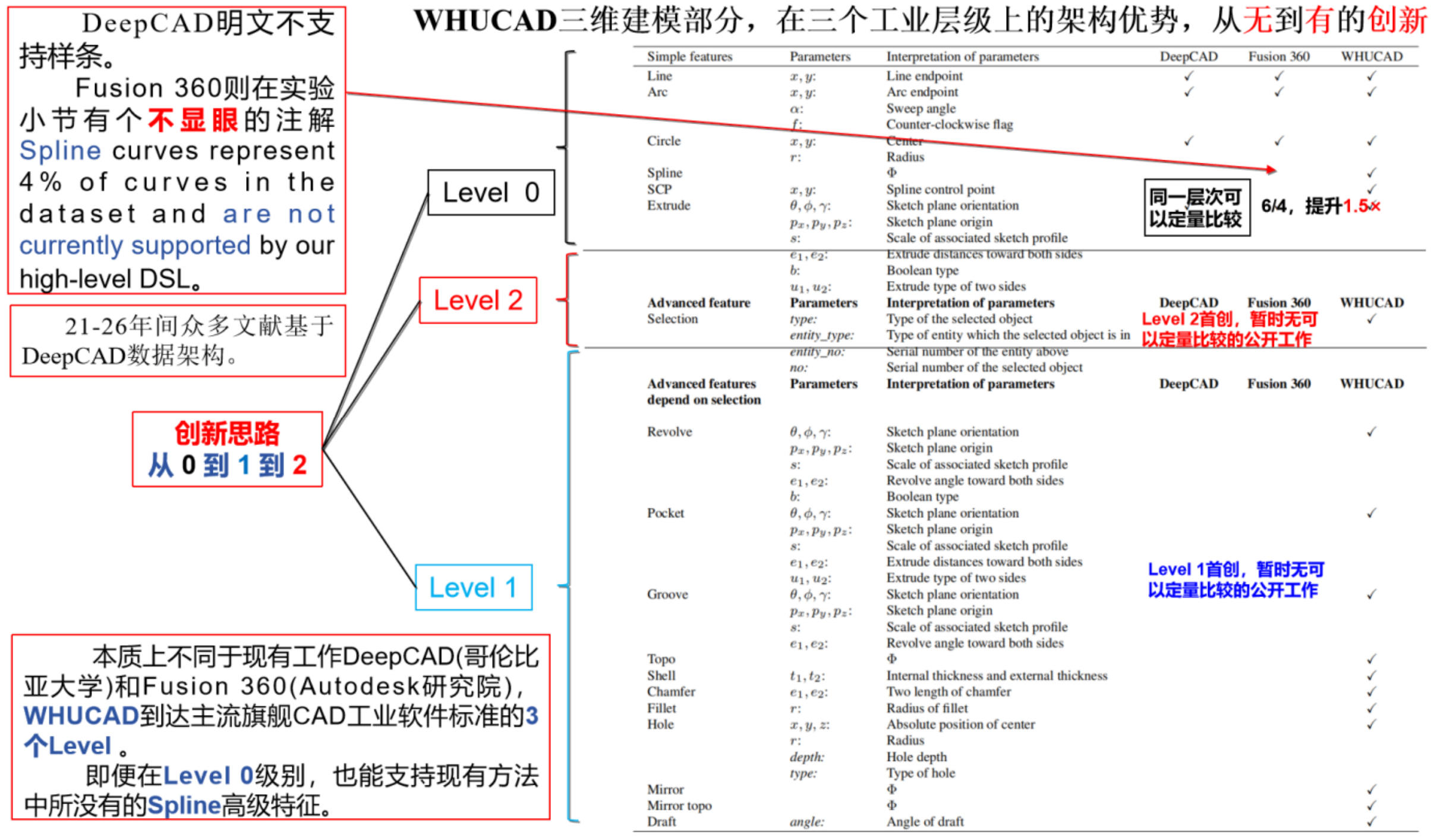

| Simple features | Parameters | Interpretation of parameters | DeepCAD | Fusion 360 | WHUCAD |
|---|---|---|---|---|---|
| Line | $x, y$: | Line endpoint | ✓ | ✓ | ✓ |
| Arc | $x, y$: | Arc endpoint | ✓ | ✓ | ✓ |
| | $\alpha$: | Sweep angle | | | |
| | $f$: | Counter-clockwise flag | | | |
| Circle | $x, y$: | Center | ✓ | ✓ | ✓ |
| | $r$: | Radius | | | |
| Spline | | Φ | | | ✓ |
| SCP | $x, y$: | Spline control point | | | ✓ |
| Extrude | $\theta, \phi, \gamma$: | Sketch plane orientation | | | |
| | $p_x, p_y, p_z$: | Sketch plane origin | | | |
| | $s$: | Scale of associated sketch profile | | | |
| | $e_1, e_2$: | Extrude distances toward both sides | | | |
| | $b$: | Boolean type | | | |
| | $u_1, u_2$: | Extrude type of two sides | | | |
| **Advanced feature** | **Parameters** | **Interpretation of parameters** | **DeepCAD** | **Fusion 360** | **WHUCAD** |
| Selection | *type:* | Type of the selected object | | | ✓ |
| | *entity_type:* | Type of entity which the selected object is in | | | |
| | *entity_no:* | Serial number of the entity above | | | |
| | *no:* | Serial number of the selected object | | | |
| **Advanced features depend on selection** | **Parameters** | **Interpretation of parameters** | **DeepCAD** | **Fusion 360** | **WHUCAD** |
| Revolve | $\theta, \phi, \gamma$: | Sketch plane orientation | | | ✓ |
| | $p_x, p_y, p_z$: | Sketch plane origin | | | |
| | $s$: | Scale of associated sketch profile | | | |
| | $e_1, e_2$: | Revolve angle toward both sides | | | |
| | $b$: | Boolean type | | | |
| Pocket | $\theta, \phi, \gamma$: | Sketch plane orientation | | | ✓ |
| | $p_x, p_y, p_z$: | Sketch plane origin | | | |
| | $s$: | Scale of associated sketch profile | | | |
| | $e_1, e_2$: | Extrude distances toward both sides | | | |
| | $u_1, u_2$: | Extrude type of two sides | | | |
| Groove | $\theta, \phi, \gamma$: | Sketch plane orientation | | | ✓ |
| | $p_x, p_y, p_z$: | Sketch plane origin | | | |
| | $s$: | Scale of associated sketch profile | | | |
| | $e_1, e_2$: | Revolve angle toward both sides | | | |
| Topo | | Φ | | | ✓ |
| Shell | $t_1, t_2$: | Internal thickness and external thickness | | | ✓ |
| Chamfer | $e_1, e_2$: | Two length of chamfer | | | ✓ |
| Fillet | $r$: | Radius of fillet | | | ✓ |
| Hole | $x, y, z$: | Absolute position of center | | | ✓ |
| | $r$: | Radius | | | |
| | *depth:* | Hole depth | | | |
| | *type:* | Type of hole | | | |
| Mirror | | Φ | | | ✓ |
| Mirror topo | | Φ | | | ✓ |
| Draft | *angle:* | Angle of draft | | | ✓ |

Figure 4. Architectural advantages of the three-dimensional modeling component of WHUCAD

### 3.3 Integrated Data Representation Architecture: Integration of WHUCAD Two-Dimensional Sketches and Three-Dimensional Modeling

The two-dimensional sketch representation and three-dimensional modeling representation of WHUCAD together form a parametric modeling representation method that moves from 2D to 3D, from primitives to features, and from geometry to historical dependencies.

- HPSketch represents advanced primitives, high-level operations, and selection mechanisms in two-dimensional sketches, providing a constrained, editable, and reusable sketch foundation for three-dimensional modeling.
- WHUCAD represents high-level features in three-dimensional modeling, selection of

three-dimensional topological elements, and parameter propagation in three-dimensional feature modeling.

Based on the above representation, the network can learn the complete native industrial-level CAD modeling process, that is, the full modeling process carried out by human engineers in an industrial-level CAD system. According to the needs of actual industrial application scenarios, two-dimensional sketches and three-dimensional modeling can be integrated.

The overall architecture of an integrated WHUCAD-pro data representation is shown in Table 1 [21].

Table 1. Data representation of WHUCAD-pro

| *Feature Category* | *HPSketch (2D)* | *WHUCAD (3D)* | *WHUCAD-pro (Integrated)* |
|---|---|---|---|
| Vector Dimensionality | 20-bit | 32-bit | 36-bit |
| Domain Focus | 2D Sketch Only | 3D Solid Only | 2D + 3D Full Loop |
| *Sketch Primitives* | | | |
| Basic (Line, Arc, Circle) | Supported | Supported | Supported |
| Advanced (Spline, Ellipse) | Supported | Not Supported | Supported |
| Conics (Hyperbola, Parabola) | Supported | Not Supported | Supported |
| *3D Features* | | | |
| Basic (Extrude, Revolve) | - | Supported | Supported |
| Engineering (Fillet, Chamfer) | - | Supported | Supported |
| Advanced (Loft, Shell, Draft) | - | Supported | Supported |
| *Constraint System* | | | |
| Geometric Constraints | Supported | Not Supported | Supported |
| Constraint Parameters | $\alpha_c, l_1, l_2, t_c$ | - | Embedded (Indices 33-36) |
| *Topology Reference* | | | |
| Selection Mechanism | Local (2D Index) | Global (B-Rep Index) | Unified Cross-Domain |

## 4 Conclusion: Two Considerations Under the Rapid Iteration of AI

Different from the goal orientation of "visual plausibility" in the sister fields of CV and CG, this paper starts from the CAD disciplinary characteristic of "industrial usability" and reviews the data representation architecture of AI+CAD. It summarizes the evolution from AI+geometric models, to AI+simple procedural models - essentially AI+solid modeling - and then to AI+industrial-level procedural models - essentially AI+parametric feature modeling.

As preliminary exploratory work, WHUCAD is the first open-source project that supports AI+industrial-level parametric feature modeling. However, it still faces the challenges brought by the rapid iteration of the AI wave, large models, and agents. Regarding how to embrace and respond to the rapid iteration of AI, two directions can be envisioned.

Direction 1: AI-assisted permanent naming, identification, and selection of topological elements.

The core component of parametric modeling is the permanent naming mechanism of topological elements and the problem of their identification and selection. The difficulty, focus, and workload of the native parametric modeling representation in WHUCAD lie in this problem. Constructing parametric models with native permanent naming of topological elements entirely by hand requires enormous effort. Therefore, introducing AI assistance is a feasible solution.

NamingCAD first creates a dataset with multimodal representations, including advanced CAD modeling processes, structured permanent naming of topological elements - modeling process plus topological information plus geometric information - boundary representation models, point cloud models, and other multimodal information. It then uses a network to predict selection results [22].

The selection results predicted by NamingCAD and CAD models with high-level features generated based on those predicted selection results are shown in Figure 5.

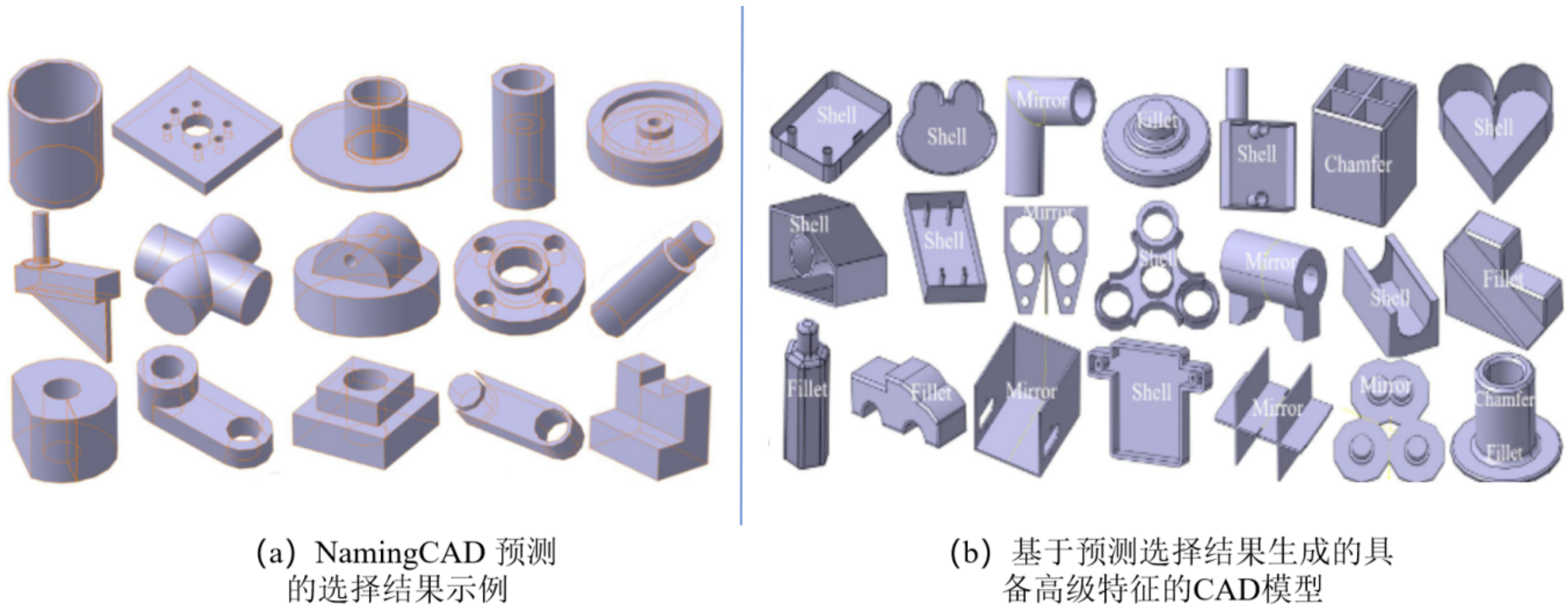

(a) NamingCAD 预测的选择结果示例

(b) 基于预测选择结果生成的具备高级特征的CAD模型

Figure 5. Visualization of the experimental results of NamingCAD

PointerCAD constructs a multimodal dataset whose selection labels contain geometric information with implicit representations, and also uses a network to predict selection results. At present, it supports chamfer and fillet [23].

Direction 2: Collaboration between large and small models, combined with agents.

The "solid modeling paradigm based on a geometry core" and the "parametric modeling paradigm based on a parametric core" can be further elevated and broken through into a new paradigm of "multi-agent-based intelligent CAD software."

The overall structure of this paradigm is shown in Figure 6.

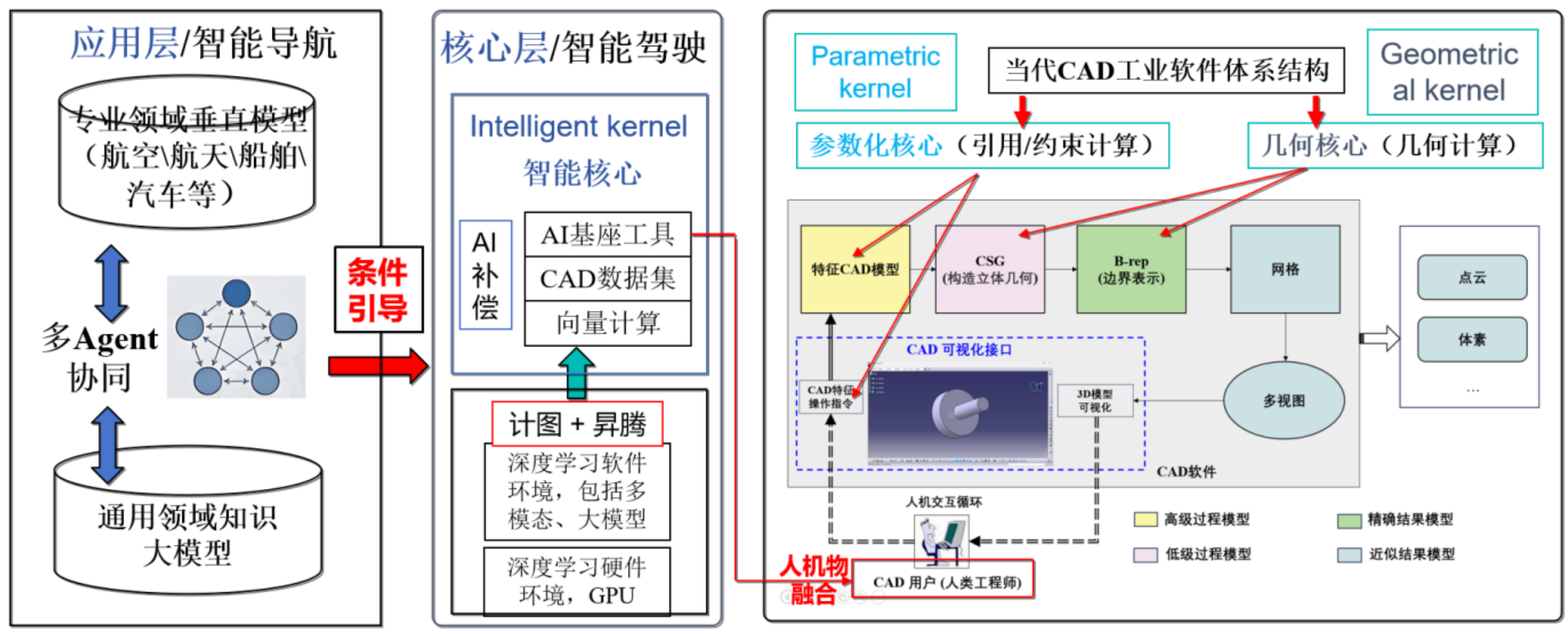

Figure 6. Overall structure of a new multi-agent-based intelligent CAD software paradigm

# AI+CAD 的数据表示架构：从 DeepCAD 的实体建模到 WHUCAD 的工业级参数化特征建模

樊儒彬 [1,2]，何发智 [1,2,3,4,5]*，刘雨辛 [2]，蔺婧 [3]，万锐博 [3]，张学成 [1]，孔庆臣 [1]

[1]武汉大学人工智能学院　武汉　430072

[2]武汉大学计算机学院/特色化示范性软件学院　武汉　430072

[3]武汉大学国家网络安全学院　武汉　430072

[4]武汉大学国家卓越工程师学院　武汉　430072

[5]武汉大学国家多媒体软件工程技术研究中心　武汉　430072

## 摘要：

2025 年 7 月中央党校主办的《学习时报》指出，95%研发设计类工业软件依赖进口，90%的高端 CAD/CAE/CAM 软件市场被欧美巨头垄断，属于典型的“卡脖子”问题。与相关领域 CV/CG 兄弟学科的“视觉可看”目标不同，CAD 学科更强调“工业可用”，数据表示架构相比网络算法优化更具基础性。首先，从 AI+CAD 的数据表示出发，汇报一种 AI+CAD 分类范式与研究进展。然后，以开源 DeepCAD 数据表示为例，分析代表性的 AI+CAD 的痛点，以及其与真实的工业级参数化特征建模的代差。接下来，对比开源 WHUCAD 数据表示，讨论其三层架构对工业级参数化特征建模的底层支撑。最后，结合 AI 浪潮的快速迭代、大模型、智能体等对 AI+工业级 CAD 进行展望。

## 引言

目前，基于深度学习的三维建模方法总体上已达到“实体建模”阶段，但距离当前工业界高端 CAD 系统（文中指是以 CATIA 为代表的卡脖子之类的高端工业 CAD 系统）

实际采用的“基于设计历史和参数化特征建模方法”仍存在明显的代差与鸿沟，成为CAD 工业软件智能化进程中的痛点。以 DeepCAD 为代表的 AI+CAD，在数据表示的底层架构上存在根本性缺陷，无论怎样扩大现有数据集的样本量，还是采用怎样先进的机器学习方法（例如大模型、多模态、Diffusion 以及 Mamba 等），都难以克服上述代差。文中以 DeepCAD 开源项目和 WHUCAD 开源项目为例，对比解读上述代差和痛点，为我国在 AI+CAD 工业软件的新赛道上，避免再次陷入“卡脖子”困境，做一份微薄的贡献。

深度学习的本质是数据表征的学习。所谓数据表征，通常是指网络模型在训练过程中从输入数据中提取和组织出的高维特征。而数据表示则更强调数据在进入网络模型之前的结构化组织方式，即研究对象被编码为何种形式、保留哪些语义关系和工程信息。对于图像、文本、离散几何等任务对象而言，数据表示相对稳定，深度学习的重点更多落在网络结构与表征学习能力的算法优化上。但是对于 CAD 而言，数据表示本身就是决定 AI 能否进入工业流程的基础问题。同一个 CAD 模型既可以表示为点云、Mesh、边界表示模型，也可以表示为草图、约束、特征树、建模次序及其拓扑命名与选择机制。不同表示方式所保留的工程语义完全不同，也直接决定了网络模型能够学习到的表征上限。因此，本文采用“数据表示”这一基础概念，重点讨论 AI+CAD 中数据对象如何被组织和表示，而不是讨论神经网络内部的特征表征及其网络算法的优化。

全文结构如下：第 1 节从 AI+CAD 的数据表示出发，汇报一种 AI+CAD 范式分类及其研究进展。第 2 节以 DeepCAD 的数据表示为例，分析 AI+简单 CAD 的痛点。第 3 节对比解读 WHUCAD 数据表示，汇报其三层数据表示架构对 AI+工业级参数化 CAD 的底层支撑。第 4 节为全文总结。

## 1 AI+CAD 范式：一种基于数据表示的分类与进展

与相关领域 CV/CG 兄弟学科的“视觉可看”目标不同，CAD 学科更强调“工业可用”，数据表示相比网络算法更具基础性。AIGC 的数据表示及生成结果的工业可用性是衡量 AI+CAD 范式的一个金标准。所以本文以 AIGC 的数据表示及生成结果，作为一种 AI+CAD 范式的分类/进展的线索，对现有 AI+CAD 研究工作进行汇报，如图 1 所示。

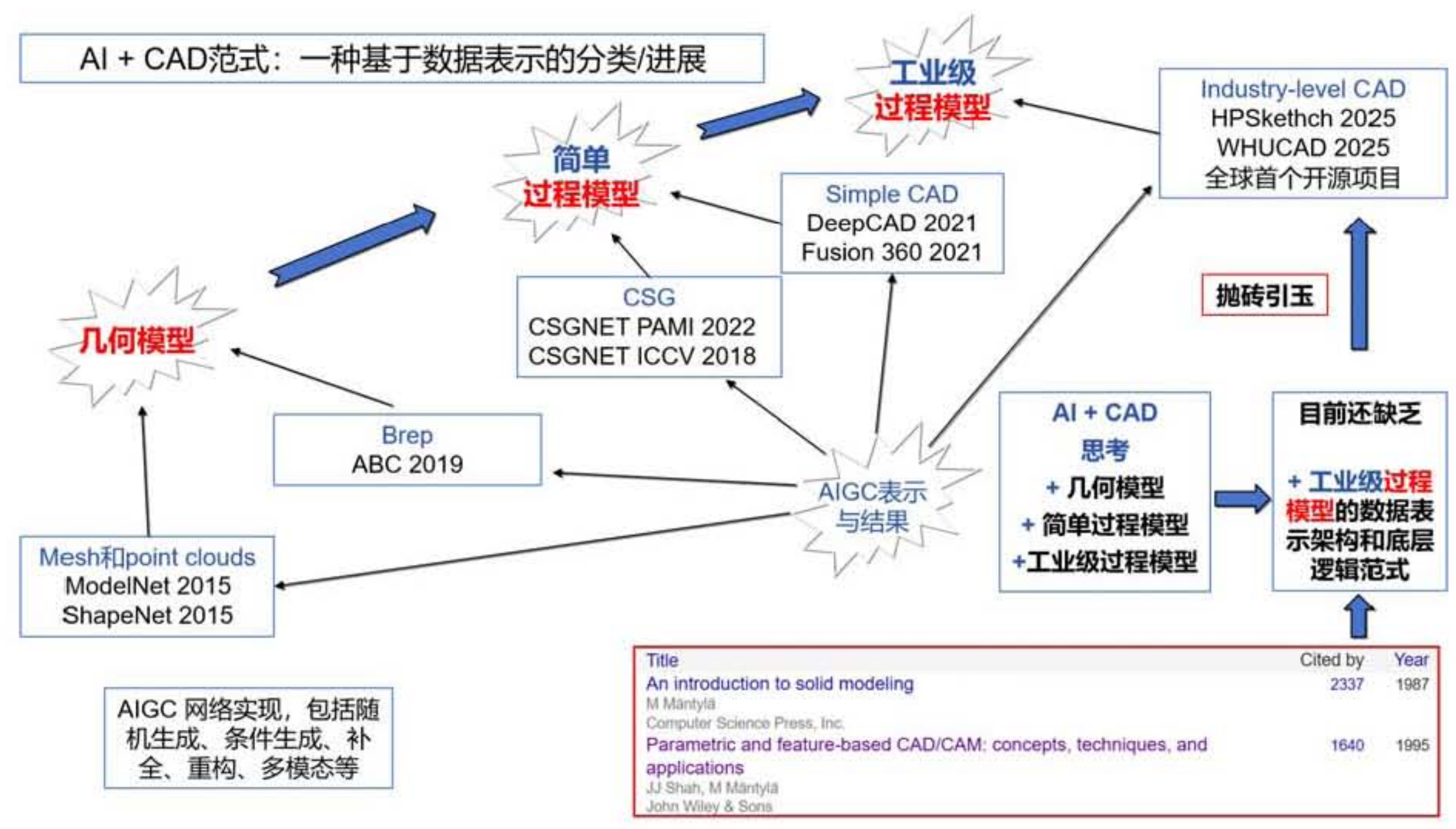

图 1. AI+CAD 范式：一种基于数据表示的分类/进展研究

## 1.1 范式 1：AI+几何表示

在计算机图形学与三维计算机视觉的深度学习研究工作中，基于三维几何的数据表示是基础。该范式以几何形状为主要学习对象，典型数据表示包括体素、点云、三角网格、隐式场以及边界表示模型。

代表性的离散、逼近表示的几何数据集 ModelNet 声称“We construct Model-Net - a large-scale 3D CAD model dataset.”但是实际只提供了三角网格。ShapeNet 则声称“It is a richly-annotated, large-scale repository of shapes represented by 3D CAD models of objects.”但是只提供点云和三角网格。这两类数据表示主要服务于三维视觉识别、重建和生成，关注的是物体外形及语义类别，而不是设计历史、约束关系以及更复杂的工业级参数化特征树[1-2]。因此，基于这类数据表示的 AIGC 系统即使能够生成视觉上合理的三维形体，也通常难以直接进入工程设计的参数修改和特征重用。

ABC 数据集在论文标题中声称是“A Big CAD Model Dataset For Geometric Deep Learning”，但在其 3.1 节进一步解释“We will use the term CAD model to refer to a geometric shape defined by a boundary representation”[3]。这类几何表示进一步把研究对象推进到连续、精确表示的边界表示模型，可以提升几何深度学习对解析曲线、曲面片、法向和曲率等几何属性的利用能力。随后，BRepGen 等工作开始尝试直接生成边界表示模型，说明 AI+CAD 研究已经从离散、逼近表示，逐步向 CAD 内核使用的连续、精

确边界表示模型发展[4]。这一类数据表示的重要贡献，在于把“视觉可看”的形状提升为“几何上更精确”的边界表示模型。基于边界表示的相关几何学习网络包括 BRepNet、ParSeNet、PrimitiveNet、SolidGen 等[5-9]，不再赘述。

基于点云、mesh、边界等几何表示的 AIGC 只能解决生成的“是什么”，而不能解决“实际建模过程中，工程师是如何一步步建模”的问题。因此，AI 直接生成几何模型，在工业可用方面至少存在两个问题。

问题一。如南佛罗里达大学教授、CAD 期刊前主编 Piegl 在 CAD 领域排名第一的十大挑战中指出：“While one can cheat the eye in computer graphics and animation, the milling machine is not as forgiving.”[10] 也就是说，在计算机图形学和动画中，几何误差可能具有视觉不可察觉性，可以“欺骗人眼”；但是在加工制造场景中，机床不会容忍几何误差。因此，AI+几何模型即使能够生成视觉上合理的三维形体，也难以直接保证其几何精度、拓扑合法性和工程可制造性满足 CAD 工业场景的需求。

问题二。则是近些年 AI+过程式模型研究动机中论述的参数化、可编辑、可重用等问题。工业 CAD 模型不仅需要描述最终几何形状，还需要保留建模历史、草图约束、参数关联和工业级的参数化特征树，以支持后续的尺寸修改、特征复用。因此，仅生成几何模型难以满足工业级参数化设计需求，不得不进一步采用过程式模型。过程式模型通过调用 CAD 几何核心/内核来输出最终几何模型，可以在一定程度上避免/缓解直接生成几何模型时的精度等问题，同时保留参数化建模过程和可编辑结构，更接近工业 CAD 软件真正需要的数据表示。

### 1.2 范式 2：AI+简单过程表示

范式 2 相对于范式 1 前进了一步，把 CAD 建模过程表示为简单的 CAD 命令序列。AI 网络学习和输出的不是显式的几何表示，而是隐式的过程表示。对于机器学习而言，这种序列化表达非常友好，便于利用 Transformer、扩散模型、自回归模型和多模态条件生成框架。

DeepCAD 是这一范式的代表，它把三维模型表示为类似自然语言的 CAD 操作 Token 序列，并基于 Transformer 进行自编码和随机生成。其公开数据集包含 178,238 个 CAD 模型及相应建模序列[11]。Fusion 360 Gallery Reconstruction 也采用类似的数据表示，并提供 8,625 条设计序列[12]。基于 AI+简单 CAD 过程模型的相关学习网络及其优化工作非常多，不再赘述[13][14]。

范式 2 中，CSG-Net 采用 CSG（Constructive Solid Geometry，构造实体几何）的数

据表示，与 DeepCAD 和 Fusion 360 的数据表示相比，在底层逻辑上是类似的[15]。DeepCAD 的命令序列，首先要通过简单的草图拉伸生成几何实体图元，然后通过一系列的布尔运算，最终得到几何结果。差别在于，CSG 是直接给出的参数化几何实体图元（例如，直接利用长、宽、高来表示一个立方体），DeepCAD 则通过草图拉伸给出参数化几何实体图元（例如，先通过一个参数化的草图四边形，加上一个拉伸的高度参数，来表示一个立方体）。两者本质上属于实体建模的范畴，仅支持图元级别的参数化。

参考 CAD 领域中具有里程碑意义的两本专著[16][17]，范式 2 本质上是 AI+实体建模（+几何内核）。

如果对标上世纪八、九十年代的工业 CAD 技术水平，这种 AI+简单过程表示（+实体建模），已经可以闭环，可以支撑实体建模，直接对接几何内核（几乎所有跟随 DeepCAD 的方法，直接对接 OCC 内核，通过全局参数的布尔运算得到几何结果）。

但是，经过多年的发展，当代工业级 CAD（文中指是以 CATIA 为代表的卡脖子之类的高端工业 CAD 系统）已经从“实体建模”发展到“基于设计历史的参数化特征建模”。所以，范式 2 无法支撑当代工业级 CAD 特征建模。

### 1.3 范式 3：AI+原生工业级过程表示

作为抛砖引玉的工作，开源项目 HPSketch 和 WHUCAD 构建了面向二维草图和三维建模的工业级 CAD 的数据表示（数据集开放下载，深度学习代码开放下载，数据基座免费使用）。与前序范式不同，WHUCAD 的工业级 CAD 过程表示，不是表示一个“形状看起来像零件”的实体，而是表示一个能够被 CAD 系统继续求解、编辑、复用和导出的参数化特征模型。因此，WHUCAD 数据表示把设计历史、草图约束、特征操作、拓扑元素命名和选择统一到同一数据表示架构中[18][19]。

因此，HPSketch 和 WHUCAD 的意义并不只是增加 CAD 命令的类型，而是把 AI+CAD 的数据表示从“简单的几何操作命令”上升到“特征操作命令”。特征是表达现代 CAD/CAM 系统中设计语义和加工工艺的一类本体概念和规则体系。

以 CAD 设计语义中倒角特征为例，尽管实体建模通过给定布尔运算的全局参数，可以得到一个倒角形状（上世纪八十年代实体建模就采用这样的方法，难以编辑和重用，实际工业部署中人工交互成本开销巨大）。但是，在现代工业级参数化特征 CAD 系统中，倒角特征的底层逻辑和本体概念如下：

- 总体上，特征既是表达工程师设计语义的一个高级 CAD 操作命令，又是特征操作的几何结果，例如边特征、面特征等。

- 就倒角特征而言，除了倒角的尺寸参数外，该高级操作需要关联一个实体边，代表了工程师的设计语义中所要修饰的几何对象（也可以称为边特征）。
- 该几何对象（边特征）又是前序一系列历史特征操作的结果。边本身就是特征，除了本身有几何形状之外（实体建模能够表示的几何形状），边特征在底层逻辑上包含了到目前为止所有的建模历史，并且一直可以追溯到第一步操作。
- 该建模历史+拓扑信息+几何信息，构建了该边特征原生的拓扑元素永久命名。这样的话，基于“边特征”的“倒角特征”才能做到真正基于设计历史的参数化特征建模（见下叙功能 2）。

不同于实体建模的单一功能，参数化特征建模有以下两个功能：

- 功能 1，通过人机交互 CAD 序列过程方式，支持当前形状的构建（这个过程等效于实体建模）。关键的差别在于，在构建当前形状的同时，同步构建各种特征，包括可视化的形状特征，以及可视化背后的特征依赖（包括原生的拓扑元素永久命名），从而支持后续功能 2。
- 功能 2，对于已经构建好的参数化特征模型，用户若编辑/修改其中的某一个特征参数，CAD 系统会根据特征依赖关系以及拓扑元素永久命名，在没有人机交互的情况下，重演建模历史，并进行特征实体的选择和定位，以及参数的自动传播，从而自动解算出调用几何内核时的布尔运算全局参数，最终得到新的几何形状。

因此，WHUCAD 的本质是 AI+工业级参数化特征建模。

## 2 DeepCAD：简单过程表示的痛点

以 DeepCAD 为例，其数据表示的是基于“简单草图”与“简单拉伸操作”的 CAD 命令操作。尽管其声称是完整的设计历史，但实际上 DeepCAD 的过程表示缺乏更底层的关键内容。相对工业级的参数化特征建模，存在明显的代差和痛点。

参考 CAD 发展史上两本里程碑式的两本专著，以及 1.3 节中倒角特征概念的举例，DeepCAD 中的简单 CAD 过程命令，本质上是简单的草图命令（二维情况）或者简单的布尔运算（三维情况），还达不到真正的参数化特征建模“特征建模”的概念。

相比当代工业级参数化特征建模系统，DeepCAD 的代差和痛点分为以下三个方面。

### 2.1 二维草图表示的不足

在二维草图方面，DeepCAD 的图元和约束表达能力与工业级 CAD 建模差距甚远。

从图元方面看，DeepCAD 仅包含直线、圆弧、圆等简单图元，难以支撑工业级 CAD 中关键的样条、抛物线、双曲线、椭圆等高级草图图元。以样条为例，CAD 学科专家学者（来自计算机、数学、机械等）经过数十年的努力，在此基础上发展的 NURBS 曲线、曲面，是现代工业级参数化 CAD 系统中形状建模的灵魂和精华。

从约束方面看，DeepCAD 的草图过程没有约束，无法支撑工业级参数化建模，也无法支撑后续的参数修改和设计意图维护。在工业级参数化 CAD 中，草图是由图元、尺寸、重合、平行、垂直、相切、对称等约束共同构成的可求解设计结构。

从二维特征依赖关系上看，由于 DeepCAD 的草图过程缺乏二维拓扑元素的命名和选择机制，无法在建模过程中建立二维图元之间的各种约束关系，从而无法支撑当代工业级 CAD 的建模。

### 2.2 缺乏特征级别的高级三维建模命令

DeepCAD 三维建模命令只是简单布尔运算，难以表达工业级 CAD 中的高级特征操作。DeepCAD 过程式表示把建模过程表达为简单草图和简单拉伸展开。多个“草图+拉伸”之间没有依赖关系。在后续编辑过程中，需要人工的方式来建立多个“草图+拉伸”之间的关系，来满足设计语义。

因此，DeepCAD 过程式表示，无法支撑旋转、倒角、圆角、镜像、拔模、抽壳、凹槽、旋转槽等高级特征操作。这些高级特征是当代工业级参数化特征建模中的基本要求。

例如，以草图和拉伸命令为例，DeepCAD 只支持垂直方向和垂直高度的拉伸。但在一个参数化特征建模中，可以拉伸到另一个指定的面，从而表达设计师的丰富设计语义。

基于拓扑元素永久命名的 CAD 命令才是真正意义上的 CAD 特征操作命名，才能支撑各种高级的参数化特征建模，包括上述“拉伸到某个面”的拉伸操作。

### 2.3 缺乏三维拓扑元素永久命名和选择机制

从三维拓扑元素永久命名和选择机制方面看，工业级参数化 CAD 系统要求特征命令能够指定其引用对象，参见 1.3 节中，现代工业级参数化特征 CAD 系统中，倒角特征的底层逻辑和本体概念。通过建模历史+拓扑信息+几何信息，构建了该边特征原生的拓扑元素永久命名，基于“边特征”的“倒角特征”才能做到真正基于设计历史的参数化特征建模。

DeepCAD 由于缺少拓扑元素永久命名和选择机制，其多个“草图+拉伸”之间没有依赖关系，同时也无法支持真正的特征级别的操作。即便在最简单的草图拉伸中，也无法支持拉伸到另一个面。

通过以上分析，DeepCAD 数据表示本质上是 AI+实体模型，其可编辑性和可重用性远远达不到当代工业级参数化特征建模的要求。

近期的相关研究表明，基于简单过程表示的 DeepCAD 数据集（17 万）和 OmniCAD 数据集（45 万），存在饱和问题（saturation issues）[20]。

## 3 WHUCAD：面向工业级参数化特征建模的数据表示架构

与 DeepCAD 简单过程表示不同，WHUCAD 的数据表示分为三级架构，总体支撑当代工业级参数化特征建模。

在 AI+CAD 时代，从 DeepCAD 到 WHUCAD 的进展，一定程度上，相当于上个世纪末传统 CAD 系统从实体建模到参数化特征建模的进展[16][17]。

### 3.1 二维草图 HPSketch 的三层级架构

WHUCAD 二维草图 HPSketch 建立了 Level 0、Level 1 和 Level 2 三层级架构，使草图不再只是“画了哪些图元”，而是在每一步的建模过程中进一步包含“对哪些图元执行了哪些操作”。其中，Level 1 和 Level 2 是首创。即使在 Level 0 层面，也支持 DeepCAD、CADL 和 SketchGraphs 中不支持的工业级高级图元。

Level 0 是图元层表示。HPSketch 在图元层创建了椭圆、样条曲线、抛物线和双曲线等工业级高级图元。这一层的意义在于提高草图的几何表达能力，使网络模型能够学习真实工业级 CAD 草图的灵魂和精华，而不是简单的线、圆、圆弧。

Level 1 是高级特征操作层表示。HPSketch 创建了镜像、倒角、圆角、旋转等高级草图数据表示。该类表示在 DeepCAD、CADL 和 SketchGraphs 中是空白。HPSketch 对标真实的工业级 CAD 草图建模。工程师并不是简单地逐条绘制图元，而是经常通过“选择已有图元—执行高级操作”的方式快速构造包括约束在内的复杂轮廓。

Level 2 是选择机制层表示，也是 HPSketch 三层级架构中的核心组件。高级操作的一个关键在于数据表示能否表达出“对哪一个图元、哪一组图元或哪一个轮廓环执行操作”。因此，HPSketch 构建了草图图元的命名规则与选择机制，使网络模型能够学习“先选定对象、再执行操作”的人类交互语义逻辑。Level 2 为 Level 1 提供引用对象，

是高级操作能够成立的前提。

HPSketch 相比 DeepCAD 的草图、CADL、SketchGraphs 的架构优势如图 2 所示。

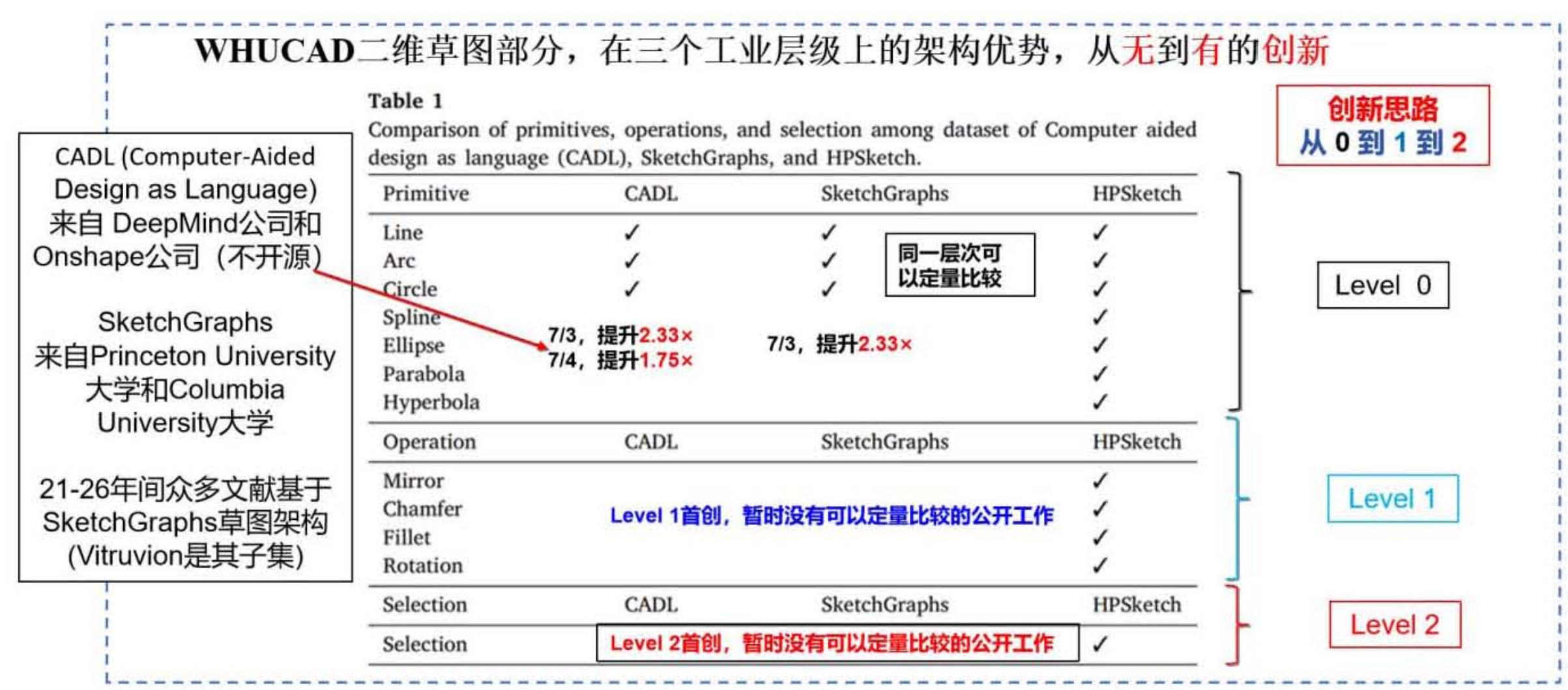

**Table 1**
Comparison of primitives, operations, and selection among dataset of Computer aided design as language (CADL), SketchGraphs, and HPSketch.

| Primitive | CADL | SketchGraphs | HPSketch |
|---|---|---|---|
| Line | ✓ | ✓ | ✓ |
| Arc | ✓ | ✓ | ✓ |
| Circle | ✓ | ✓ | ✓ |
| Spline | | | ✓ |
| Ellipse | | | ✓ |
| Parabola | | | ✓ |
| Hyperbola | | | ✓ |
| **Operation** | **CADL** | **SketchGraphs** | **HPSketch** |
| Mirror | | | ✓ |
| Chamfer | | | ✓ |
| Fillet | | | ✓ |
| Rotation | | | ✓ |
| **Selection** | **CADL** | **SketchGraphs** | **HPSketch** |
| Selection | | | ✓ |

图 2. WHUCAD 二维草图部分的架构优势

架构图之外需要进一步的分析和说明如下。

第一，SketchGraphs 的草图实际上是图表示的“图元+约束”结果（a geometric constraint graph），不支持过程建模。基于该表示，网络只能学习到该结果，学不到过程和原因，属于典型的“知其然，不知其所以然”。相比之下，HPSketch 支持过程建模，不仅“知其然”，还“知其所以然”。

第二，HPSketch 的约束是基于“过程+选择”的，上述架构图之外的约束如图 3 所示，总体覆盖了主流旗舰 CAD 参数化建模系统的草图约束类型。

| Constraint | ......(the former digits) | $\alpha_c$ | $l_1$ | $l_2$ | $t_c$ |
|---|---|---|---|---|---|
| Fix | | | | | 0 |
| Distance | | | | $l_2$ | 1 |
| Coincidence | | | | | 2 |
| Concentricity | | | | | 3 |
| Tangency | | | | | 4 |
| Length | | | | $l_2$ | 5 |
| Angle | | $\alpha_c$ | | | 6 |
| Parallelism | | | | | 7 |
| Horizontality | | | | | 8 |
| Perpendicularity | | | | | 9 |
| Verticality | | | | | 10 |
| Radius | | | $l_1$ | | 11 |
| Symmetry | | | | | 12 |
| Equidistance | | | | | 13 |
| MajorRadius | | | $l_1$ | | 14 |
| MinorRadius | | | $l_1$ | | 15 |

图 3. HPSketch 的约束表示

论文已投稿至《计算机辅助设计与图形学学报》，尚处于审稿环节。

### 3.2 三维建模 WHUCAD 的三层级架构

WHUCAD 三维建模部分建立了 Level 0、Level 1 和 Level 2 三层级架构。其中，Level 1 和 Level 2 是首创。即使在 Level 0 层面，也支持 DeepCAD 和 Fusion 360 中所没有的样条高级图元。

在 Level 1 层面，WHUCAD 在三维建模表示中创建旋转、凹槽、旋转槽、抽壳、倒角、圆角、孔、镜像、拔模等高级特征。其 Level 1 层面的数据表示，总体覆盖了主流旗舰 CAD 参数化建模系统的高级特征建模能力。

在 Level 2 层面，WHUCAD 创建了原生的三维拓扑元素永久命名与选择机制。三维特征建模中的选择对象可能是由前序操作生成的边或面。例如，倒角操作必须明确作用于哪一条边，抽壳操作必须明确移除或保留哪一个面。WHUCAD 具备拓扑命名与选择机制，建立前后不同特征操作之间的引用关系，从而支撑设计历史变动时的参数传播和特征重建。一旦前序的特征操作变化，可以自动维护设计语义。WHUCAD 三维建模部分的架构优势如图 4 所示。

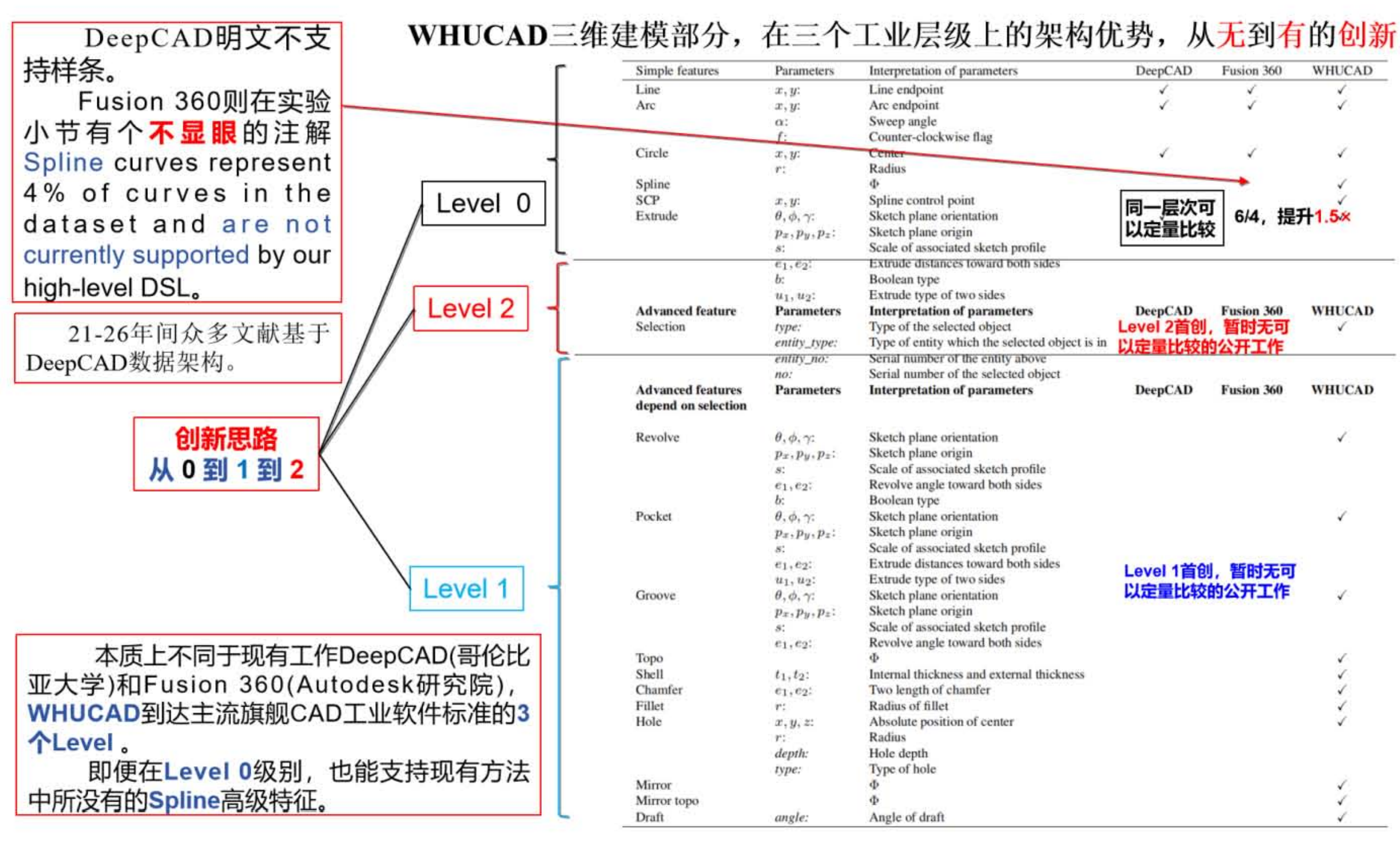

图 4. WHUCAD 三维建模部分的架构优势

### 3.3 一体化数据表示架构：WHUCAD 二维草图和三维建模的集成

WHUCAD 的二维草图表示和三维建模表示共同构成了从二维到三维、从图元到特征、从几何到历史依赖的参数化建模表示方法。

论文已投稿至《计算机辅助设计与图形学学报》，尚处于审稿环节。

- HPSketch 表示了二维草图中的高级图元、高级操作和选择机制问题，为三维建模提供可约束、可编辑、可复用的草图基础。
- WHUCAD 表示了三维建模中的高级特征、三维拓扑元素的选择和三维特征建模的参数传播问题。

基于上述表示，网络能够学习到原生的工业级别的 CAD 建模的全过程，即人类工程师在工业级 CAD 系统上完整的建模过程。根据实际工业应用场景的需求，可以把二维草图和三维建模集成在一起。

一个集成化的 WHUCAD-pro 的数据表示总体架构[21]，如表 1 所示。

表 1. WHUCAD-pro 的数据表示

| *Feature Category* | *HPSketch (2D)* | *WHUCAD (3D)* | *WHUCAD-pro (Integrated)* |
|---|---|---|---|
| Vector Dimensionality | 20-bit | 32-bit | 36-bit |
| Domain Focus | 2D Sketch Only | 3D Solid Only | 2D + 3D Full Loop |
| *Sketch Primitives* | | | |
| Basic (Line, Arc, Circle) | Supported | Supported | Supported |
| Advanced (Spline, Ellipse) | Supported | Not Supported | Supported |
| Conics (Hyperbola, Parabola) | Supported | Not Supported | Supported |
| *3D Features* | | | |
| Basic (Extrude, Revolve) | - | Supported | Supported |
| Engineering (Fillet, Chamfer) | - | Supported | Supported |
| Advanced (Loft, Shell, Draft) | - | Supported | Supported |
| *Constraint System* | | | |
| Geometric Constraints | Supported | Not Supported | Supported |
| Constraint Parameters | $\alpha_c, l_1, l_2, t_c$ | - | Embedded (Indices 33-36) |
| *Topology Reference* | | | |
| Selection Mechanism | Local (2D Index) | Global (B-Rep Index) | Unified Cross-Domain |

## 4 结束语：AI 快速迭代浪潮下的两点考虑

不同于本领域 CV/CG 兄弟方向“视觉可看”的目标导向，文中从“工业可用”这一 CAD 学科特点出发，综述了 AI+CAD 数据表示架构，从 AI+几何模型，到 AI+简单过程模型（本质上是 AI+实体建模），再到 AI+工业级过程模型（本质上是 AI+参数化特征建模）的演进。

WHUCAD 作为抛砖引玉的工作，是首个支撑 AI+工业级参数化特征建模的开源项目。但是仍然面临 AI 浪潮迅速迭代、大模型和智能体的挑战。如何拥抱和迎接 AI 的浪潮快速迭代，可以展望两个方向。

方向一，AI 辅助拓扑元素永久命名和辨识选择。

参数化建模的核心组件是拓扑元素的永久命名机制及其辨识选择问题。WHUCAD的原生参数化建模表示的难点、重点和工作量在于该问题。完全基于人工来构建具有原生拓扑元素的永久命名机制的参数化模型的工作量巨大，因此引入 AI 辅助是一个可行的方案。

文献 NamingCAD，首先创建了一个多模态表示的数据集，包含高级 CAD 建模过程、结构化拓扑元素永久命名（建模过程+拓扑信息+几何信息）、边界表示模型和点云模型等多模态信息，然后用网络来预测选择结果[22]。

NamingCAD 预测的选择结果及其基于预测选择结果生成的具备高级特征的 CAD 模型如图 5 所示。

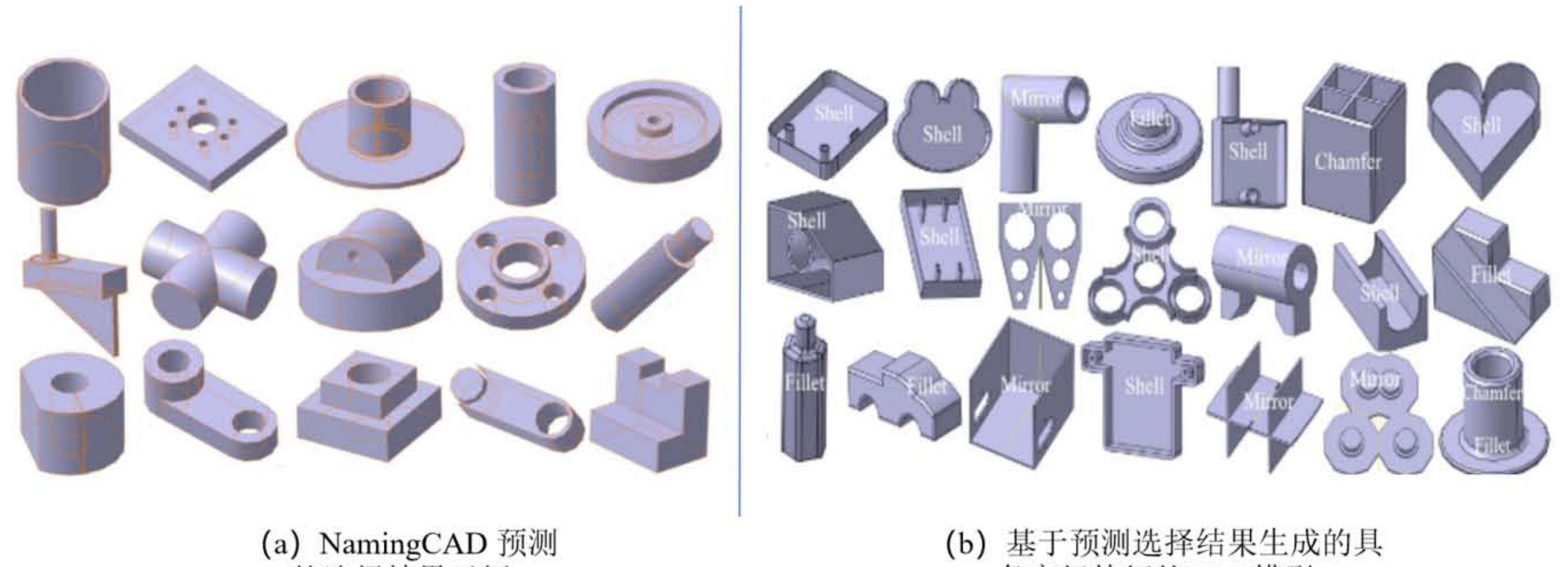

(a) NamingCAD 预测的选择结果示例

(b) 基于预测选择结果生成的具备高级特征的CAD模型

图 5. NamingCAD 的实验结果可视化

文献 PointerCAD 构建了带有隐式表示的几何信息作为选择标签的多模态数据集，也用网络预测选择结果，目前支持倒角和圆角[23]。

方向二，大小模型协同，结合智能体。

把“基于几何核心的实体建模范型”和“基于参数化核心的参数化建模范型”进一步升华，突破到“基于多 Agent 的智能化 CAD 软件新范型”。

该范型的总体结构如图 6 所示。

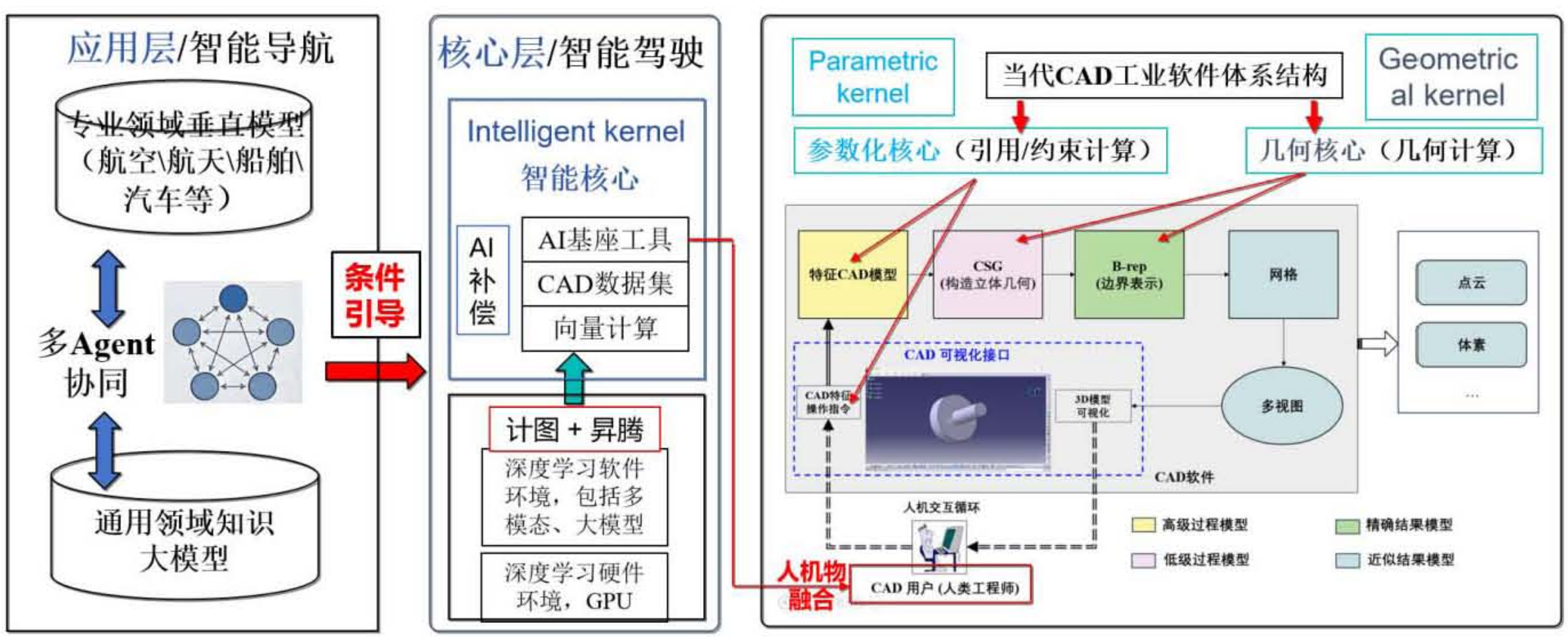

图 6. 基于多 Agent 的智能化 CAD 软件新范型总体结构

## 参考文献